\documentclass[a4paper,fleqn,usenatbib]{mnras}

\hypersetup{linkcolor=red,citecolor=blue,filecolor=cyan,urlcolor=magenta}

\usepackage{threeparttable}
\usepackage{xcolor}
\definecolor{andrewcolor}{RGB}{50,205,50}
\definecolor{samcolor}{RGB}{255,10,25}
\definecolor{sarahcolor}{RGB}{167, 66, 244}

\usepackage{newtxtext,newtxmath}

\usepackage[T1]{fontenc}
\usepackage{ae,aecompl}

\usepackage{graphicx}	
\usepackage{amsmath}	

\title[GMC lifetimes in FIRE-2 simulations]{Live Fast, Die Young: GMC lifetimes in the FIRE cosmological simulations of Milky Way-mass galaxies}

\author[S. M. Benincasa et al.]{
Samantha M. Benincasa,$^{1}$\thanks{E-mail: smbenincasa@ucdavis.edu}
Sarah R. Loebman, $^{1}$\thanks{Hubble Fellow}
Andrew Wetzel,$^{1}$ Philip F. Hopkins,$^{2}$
\newauthor 
Norman Murray,$^{3,4}$ Matthew A. Bellardini,$^{1}$ Claude-Andr\'e Faucher-Gigu\`ere,$^{5}$ 
\newauthor D\'avid Guszejnov,$^{6}$ and Matthew Orr$^{2}$
\\
$^{1}$Department of Physics, University of California, Davis, CA 95616, USA\\
$^{2}$TAPIR, Mailcode 350-17, California Institute of Technology, Pasadena, CA, 91125\\
$^{3}$Canadian Institute for Theoretical Astrophysics, 60 St. George Street, University of Toronto, ON M5S 3H8, Canada\\
$^{4}$ Canada Research Chair in Astrophysics\\
$^{5}$Department of Physics and Astronomy and Center for Interdisciplinary Exploration and Research in Astrophysics (CIERA), Northwestern University, \\2145 Sheridan Road, Evanston, IL 60208, USA 78712, USA\\
$^{6}$Department of Astronomy, The University of Texas at Austin, Austin, TX\\
}

\date{Accepted XXX. Received YYY; in original form ZZZ}

\pubyear{2019}

\begin{document}
\label{firstpage}
\pagerange{\pageref{firstpage}--\pageref{lastpage}}
\maketitle

\begin{abstract}
We present the first measurement of the lifetimes of Giant Molecular Clouds (GMCs) in cosmological simulations at $z = 0$, using the Latte suite of FIRE-2 simulations of Milky Way-mass galaxies. We track GMCs with total gas mass $\gtrsim 10^5$ M$_\odot$ at high spatial ($\sim1$ pc), mass ($7100$ M$_{\odot}$), and temporal (1 Myr) resolution. Our simulated GMCs are consistent with the distribution of masses for massive GMCs in the Milky Way and nearby galaxies. We find GMC lifetimes of $5-7$ Myr, or 1-2 freefall times, on average, with less than 2\% of clouds living longer than 20 Myr. We find decreasing GMC lifetimes with increasing virial parameter, and weakly increasing GMC lifetimes with galactocentric radius, implying that environment affects the evolutionary cycle of GMCs. However, our GMC lifetimes show no systematic dependence on GMC mass or amount of star formation. These results are broadly consistent with inferences from the literature and provide an initial investigation into ultimately understanding the physical processes that govern GMC lifetimes in a cosmological setting.

\end{abstract}




\section{Introduction}

\begin{figure*}
	\includegraphics[width=\textwidth]{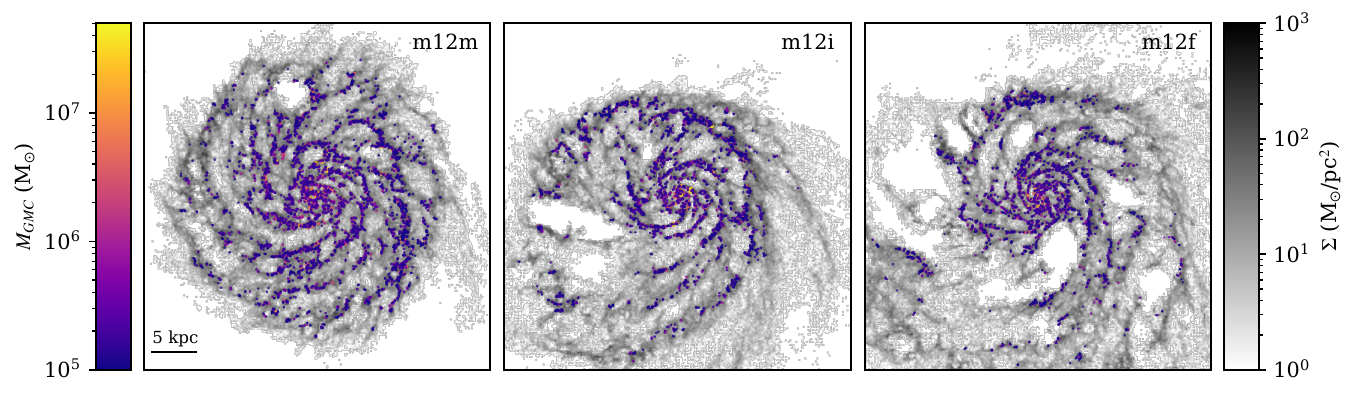}
	\caption{Maps of gas surface density at $z = 0$ for the 3 cosmological zoom-in simulations in this work. We show the cold gas surface density via the greyscale map in the background. We overlay the locations of GMCs, coloured by their mass. Each image is 40 kpc across.
	At a given time, we identify $\sim1200$ GMCs in {\tt m12i}, $\sim2700$ GMCs in {\tt m12m} and $\sim1600$ GMCs in {\tt m12f} with $M \gtrsim 10^5$ M$_{\odot}$.}
	\label{fig:faceon}
\end{figure*}

As the birthplace of stars, GMCs are fundamental to our understanding of star formation and the baryon life cycle. Because we are entering an era of high-precision measurements of dense gas in the Milky Way (MW) and nearby galaxies, we are newly positioned to make great strides in our understanding of the physics governing GMCs \citep[e.g.][]{schinnerer2019}.

The galactic environment, including the dynamical state of the ISM, affects the properties of GMCs: there is no universal set of cloud properties across all galaxies.
For example, \citet{sun2018} find that, for GMCs with a small range in virial parameters, different galactic environments can drive wildly different internal states, such as turbulent pressure. Furthermore, understanding pressure confinement in the ISM may be important to understanding the internal states of GMCs \citep[e.g.][]{faesi2018, schruba2019}.

We only now are beginning to understand the connections between star formation and the life cycle of GMCs. Constraining the lifetimes of GMCs is critical to constraining the physics of the cycle of star formation, including how dense gas cycles through the ISM. The amount of time gas spends in the star-forming state can explain the long depletion time in galaxies \citep{semenov2017}, although larger-scale galactic equilibria ultimately may determine the low efficiency of galactic star formation \citep[e.g.][]{ostriker2010}. In principle, GMC lifetimes are sensitive to the form(s) of stellar feedback that are most critical to truncating star formation \citep[e.g.][]{lopez2014, howard2017, kruijssen2019}.

Recent work, both theoretical and observational, has shed light on the lifetimes of GMCs. Observational measurements of lifetimes require a statistical approach, because we cannot track GMCs in real time. \citet{miura2012} use the connection between the evolutionary states of GMCs and young stellar objects to infer lifetimes in M33 of $20 - 40$ Myr. Another approach is to compare the spatial distributions of gas tracer peaks and stellar tracer peaks \citep{schruba2010, kruijssen2014, kruijssen2018}. \citet{kruijssen2019} infer lifetimes of $\sim 10$ Myr in NGC 300, based on the lifetime of the CO emitting gas in the clouds. In the MW, \citet{murray2011} estimate lifetimes of massive GMCs of $27 \pm 12$ Myr. However, these are indirect inferences of GMC lifetimes.

A large catalog of theoretical work now compliments these observational studies. Analytical calculations favour GMC lifetimes of a few crossing/dynamical times \citep[e.g.][]{krumholz2006, elmegreen2007}. Recently, \citet{jeffreson2018} have built upon this work, comparing important physical timescales impacting GMC evolution, leading to an estimated GMC lifetime of $10 - 50$ Myr. Furthermore, various works have used isolated (non-cosmological) simulations to achieve the necessary high dynamic range across a full galactic disk to disentangle environmental dependence \citep[e.g.][]{ward2016, grisdale2018,pettitt2019, dobbs2019}. Here, one can track the evolution of single cloud or cloud complex. Highly resolved studies of individual GMCs have found lifetimes of typically $\sim1-2$ freefall times \citep[e.g.][]{harper2011, grudic2018}. However, these works do not capture the galactic environment, global evolution, confinement from the surrounding ISM nor accretion.
In isolated (non-cosmological) galaxy simulations, one can measure GMC lifetimes by tracking mass gain and loss of a full population of GMCs over time \citep{hopkins2012, dobbs2013, grisdale2019}. For example, \citet{dobbs2013} find cloud lifetimes of $4-25$ Myr in isolated (non-cosmological) galaxy simulations with imposed spiral potentials. Similarly, \citet{hopkins2012} find cloud lifetimes with a median of $4-5$ Myr.

Cosmological galaxy simulations now offer new laboratories for examining GMC properties and lifetimes directly in galactic environments in cosmological settings. Suites of cosmological zoom-in simulations now offer sufficient dynamic range across a range of galactic morphologies and properties, including larger-scale processes like cosmic gas accretion, wind recycling, and perturbations from satellite galaxies. These are important sources for driving turbulence in the interstellar medium (ISM).

In this paper, we present the first measurement of GMC lifetimes in cosmological zoom-in simulations of MW/M31-mass galaxies at $z\!=\!0$, examining dependence on galactic environment, GMC mass, and GMC star-formation activity. In section \ref{sec:sims} we describe the Latte suite of FIRE-2 simulations that we employ in this work, as well as our cloud tracking algorithm.

\section{Methods} \label{sec:sims}
We analyze 3 galaxies from the Latte suite of FIRE-2 cosmological zoom-in simulations of MW/M31-mass galaxies \citep{wetzel2016}. We ran these simulations using the FIRE-2 physics model \citep{hopkins2018}, employing the Lagrangian meshless finite-mass hydrodynamics code \textsc{Gizmo} \citep{hopkins2015}. These simulations explicitly model stellar feedback from core-collapse and type Ia supernovae, stellar winds, photoionization, photoelectric heating and radiation pressure, as detailed in \citet{hopkins2018}, including gas heating and cooling across $10-10^{10}$ K. Star formation occurs in gas that is self-gravitating, Jeans-unstable, cold ($T < 10^4$ K), dense ($n > 1000$ cm$^{-3}$), and molecular \citep[following][]{krumholz2011}. These simulation have gas and (intial) star particle masses of 7100 M$_{\odot}$. Gas hydrodynamic smoothing is fully adaptive and is identical to force softening, reaching a minimum of 1 pc (Plummer equivalent), with force softening in the typical ISM (densities $\sim1$ cm$^{-3}$) of $\sim20$ pc. The force softening of star and dark-matter particles is 4 and 40 pc.

We focus on 3 galaxies that are particularly MW/M31-like in mass and size: {\tt m12i}, {\tt m12m}, and {\tt m12f} \citep{wetzel2016, hopkins2018, sanderson2018}. These span a range in morphology: {\tt m12m} is a flocculent spiral while {\tt m12f} has had a recent interaction resulting in a slightly disturbed morphology. The total stellar masses for {\tt m12m}, {\tt m12i} and {\tt m12f} are $7.9\times10^{10}$ M$_{\odot}$, $6.3\times10^{10}$ M$_{\odot}$ and $5.1\times10^{10}$ M$_{\odot}$, respectively. The total gas masses for {\tt m12m}, {\tt m12i} and {\tt m12f} are $2.1\times10^{10}$ M$_{\odot}$, $1.6\times10^{10}$ M$_{\odot}$ and $2.3\times10^{10}$ M$_{\odot}$, respectively as measured within $R_{90}$.  For comparison, as measured within the virial radius, the total baryonic mass of the MW is $8.5\pm 1.3 \times 10^{10}$ M$_{\odot}$, with $5.1\times10^{10}$ M$_{\odot}$ in stars \citep{blandhawthorn2016}.
For this work, we re-simulated these 3 galaxies to store snapshots every 1 Myr over the final 100 Myr before $z=0$.
Figure~\ref{fig:faceon} shows gas surface density maps for these 3 galaxies at $z\!=\!0$.
Several works have examined the ISM properties of these simulated galaxies \citep{sanderson2018, orr2018, elbadry2018a, elbadry2018b, hung2019, guszejnov2019}.

\begin{figure}
    \centering
    \includegraphics[width=0.99\columnwidth]{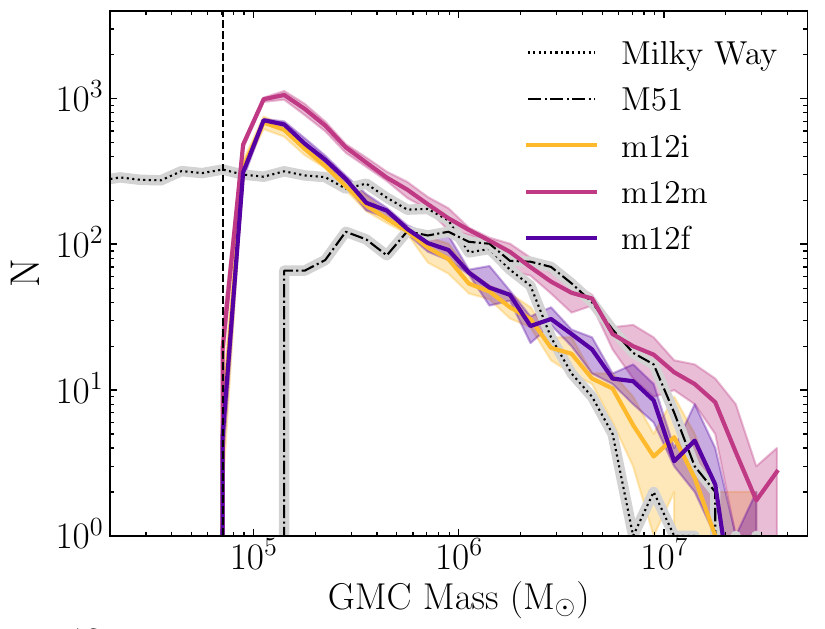}
    \caption{The distribution of GMC masses in our 3 cosmological simulations at $z\!\approx\!0$. The shaded region shows the range across different snapshots.
    We compare to observed GMCs in the Milky Way \citep{miville2017} and M51 \citep{colombo2014}. Our simulated GMC mass distributions are broadly within the ranges of those observed, though we note an apparent excess at low mass, near our resolution limit (20 gas elements).
    }
    \label{fig:obs}
\end{figure}

\begin{figure}
    \centering
    \includegraphics[width=0.99\columnwidth]{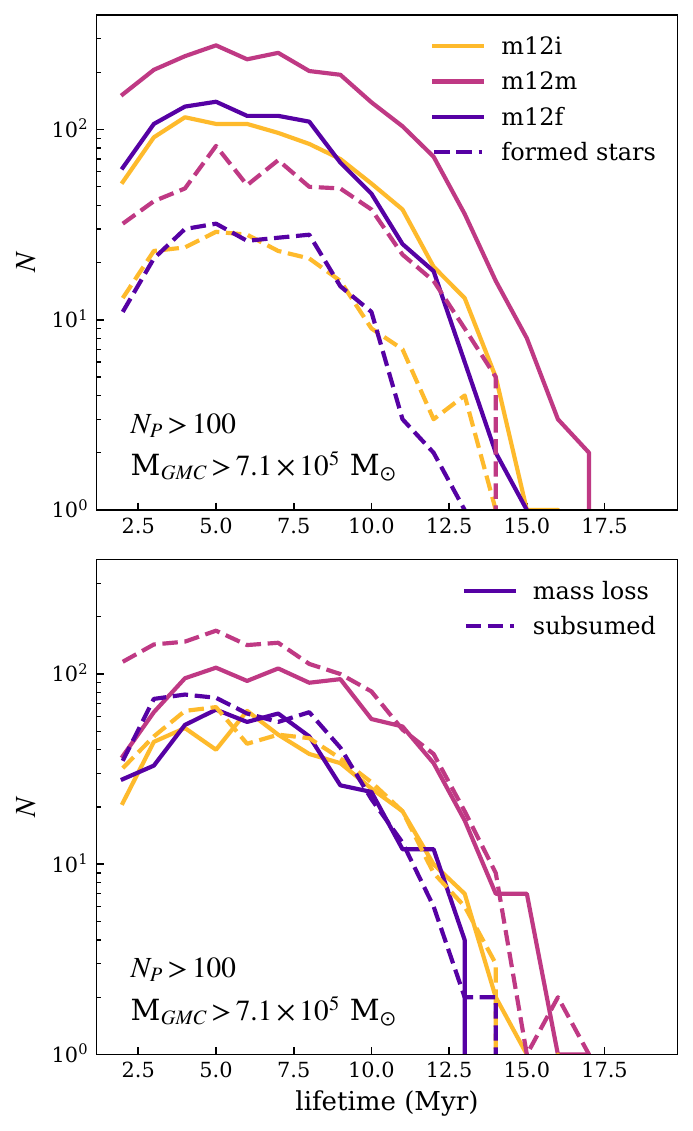}
    \caption{The distribution of GMC lifetimes in each of our 3 cosmological simulations at $z\!\approx\!0$, for clouds with at least 100 resolution elements. The mean lifetime is $\sim6$ Myr, with only 1.7\% of clouds living longer than 20 Myr. Dashed curves show clouds that hosted any level of star formation; we do not find significant differences based on star-formation activity.
    These lifetimes are broadly consistent with both observational and theoretical work (typically $10-20$ Myr); though our average favours shorter lifetimes, in the massive GMCs that we resolve. In the bottom panel we divide the GMCs into those that died from mass loss (solid line) and those that died from being subsumed into a more massive cloud (dashed line). Subsumed clouds show slightly shorter lifetimes, particularly in {\tt m12m}. However, this bias is small and the peak of the lifetime distribution remains largely unaffected.}
    \label{fig:life}
\end{figure}

\subsection{Identifying Clouds}
We identify GMCs using a Friends-of-Friends (FoF) algorithm, which groups gas elements based on proximity via an isodensity threshold. FoF requires a single free parameter, the linking length, $l$, which determines the isodensity threshold. After extensive testing, we choose a linking length of 20 pc for identifying GMCs, which corresponds to a local density of $\sim30$ cm$^{-3}$. Additionally, we consider only gas elements with hydrodynamic kernel densities $>10$ cm$^{-3}$ and with temperatures below $10^4$ K. We confirmed that these cuts have no impact on the identified GMCs, but they significantly increases the speed of finding. This method is consistent with that of Lakhlani et al. (in prep). Figure~\ref{fig:faceon} shows the locations of these GMCs, overlaid on the maps of total gas surface density. We analyze GMCS with more than 20 gas elements, corresponding to a minimum mass of $\sim 10^5$ M$_{\odot}$. Thus, by GMC `mass' we mean the sum of the masses of all gas elements in the cloud.

Figure~\ref{fig:obs} shows the properties of GMCs in our simulations at $z\!=\!0$, compared with GMCs observed in both the MW and M51. For comparison, M51 has a stellar mass of $3.6\times 10^{10}$ M$_{\odot}$ and a gas fraction of 0.2 \citep{shetty2007, leroy2008, schinnerer2013}. The MW has a stellar mass of $6\times 10^{10}$ M$_{\odot}$ \citep{licquia2015}. The top panel shows the mass distribution of GMCs in our simulated galaxies compared to GMCs in M51 (dot-dashed line) and the MW (dotted line). We use the M51 sample from \citet{colombo2014} and the MW sample from \citet{miville2017}. The rollover of the GMC mass distribution for M51 shows the position of the adopted completeness limit of $3.6\times10^5$ M$_{\odot}$ \citep{colombo2014}. In contrast, studies of the MW are complete down to lower masses \citep{miville2017, romanduval2010}. The vertical dashed line shows our resolution limit, which matches closely the completeness limit for M51. From this comparison it appears that our simulations produce a possible excess of GMCs approaching the mass resolution limit. However, the shape of the mass function at the resolution limit is particularly sensitive to the GMC identification algorithm. Further, while we currently probe only the massive GMCs in our simulations, in future work we will resolve an order of magnitude lower cloud mass.
The virial parameter is defined as
\begin{equation}
\alpha_{\text{vir}} = \frac{2E_{\text{kin}}}{|E_{\text{grav}}|} = \frac{5GM}{\sigma_v^2 R}
\end{equation}
where $E_{\text{kin}}$ is the kinetic energy of the cloud, $E_{\text{grav}}$ is the gravitational potential energy of the cloud, M is the mass of the cloud, and R its radius assuming a spherical shape. The virial parameter is meant to assess the balance of gravitational and kinetic energy in a GMC. In our case, since we have detailed knowledge of the gas elements that constitute the clouds, we directly calculate the kinetic and gravitational potential energy of each cloud and calculate $\alpha_{\text{vir}}$ in this way. Our GMCs have a large range of virial parameters (boundedness), with the distribution peaking at $\alpha_{vir}$ between 2.5 and 3.5.

\subsection{Tracking the Evolution of Clouds}\label{sec:track}

To track the evolution of clouds, we follow the methodology used in previous studies \citep[e.g.][]{hopkins2012, dobbs2013}. Given a specific GMC $i$ in snapshot $n$, we identify the descendant or progenitor of this GMC (in snapshot $n \pm 1$) as the FOF group that contains the most total mass in elements from the original GMC $i$. We take the reference GMC at arbitrary snapshots, spaced by 10 Myr to avoid double counting.
We track each cloud forward and backward until it has lost 50\% of its original/reference mass at the midpoint snapshot. \citet{richings2016} have shown the measured lifetime can vary depending on what subset of the GMC is used to assess this fraction. In this work, this mass evaluation is made irrespective of whether or not the particles were part of the reference mass. We have performed a comparison of the methods discussed in \citet{richings2016} in our GMC population for {\tt m12i}. We find a difference in the median GMC lifetime of $\sim 1$ Myr.

We also stop tracking if the cloud's elements no longer make up the main constituent after a merger, for example, when a cloud is absorbed/subsumed into a larger cloud. This subsumed population, across all hosts, constitutes $\sim69\%$ of all clouds, and $\sim 57\%$ of clouds in our population of most massive clouds, with $N_p\!>\!100$. In the following sections of the paper we present both comparisons of cloud properties and comparisons of the subsumed population with the mass loss cloud population.

Of course, the choice of mass threshold impacts the resultant lifetime measurement. We tested this by examining lifetimes using mass cutoffs of $1/2$, $1/e$, and $1/5$ of the original mass. These lead to small changes to the peak of the GMC lifetime distribution, increasing it by $1-2$ Myr. Appendix \ref{sec:app} presents these differences.
Further, we note that we measure lifetimes of the overdense gas cloud as a coherent unit; we do not follow the creation of destruction of molecular gas, so are {\em not} measuring the lifetimes of molecules or other species within clouds.
One should note these caveats in making comparisons to observationally measured GMC lifetimes.

\section{Results} \label{sec:results}

\subsection{The distribution of GMC lifetimes}

We now present our first results on GMC lifetimes via cloud tracking. To increase the number of GMCs in our sample, we stack the results from multiple snapshots in the 100 Myr preceding $z\!=\!0$. To prevent double-counting, we space the reference snapshots to be 10 Myr apart, comparable to the longest GMC lifetimes that we find.
Figure~\ref{fig:life} shows the distribution of GMC lifetimes. We include only GMCs beyond galactocentric radius of 1 kpc,
and we examine only our most resolved clouds, with more than 100 elements. Generally, we find a mean lifetime of $\sim 6$ Myr, with only 1.7\% of all clouds having lifetimes longer than 20 Myr. Table~\ref{tab:life} lists these mean lifetimes with their 1-sigma scatter. In the top panel solid lines show all GMCs, while dashed lines show only those that formed stars during their lifetime.

In the bottom panel of Figure \ref{fig:life} we divide the sample of GMCs by the way that they are destroyed. This can occur either by a mass loss of 50\% from their reference mass (solid line) or by being subsumed by a larger cloud (dashed line). We find that clouds that are subsumed prefer slightly shorter lifetimes, particularly in {\tt m12m}. However, this generally leaves the peak of the lifetime distributions unchanged. Again, Table \ref{tab:life} lists the averages and standard deviations for each case. Overall, we find only weak difference between clouds that are destroyed from mass loss and clouds that are subsumed or undergo collisions.

\begin{figure*}
    \centering
    \includegraphics[width=0.9\textwidth]{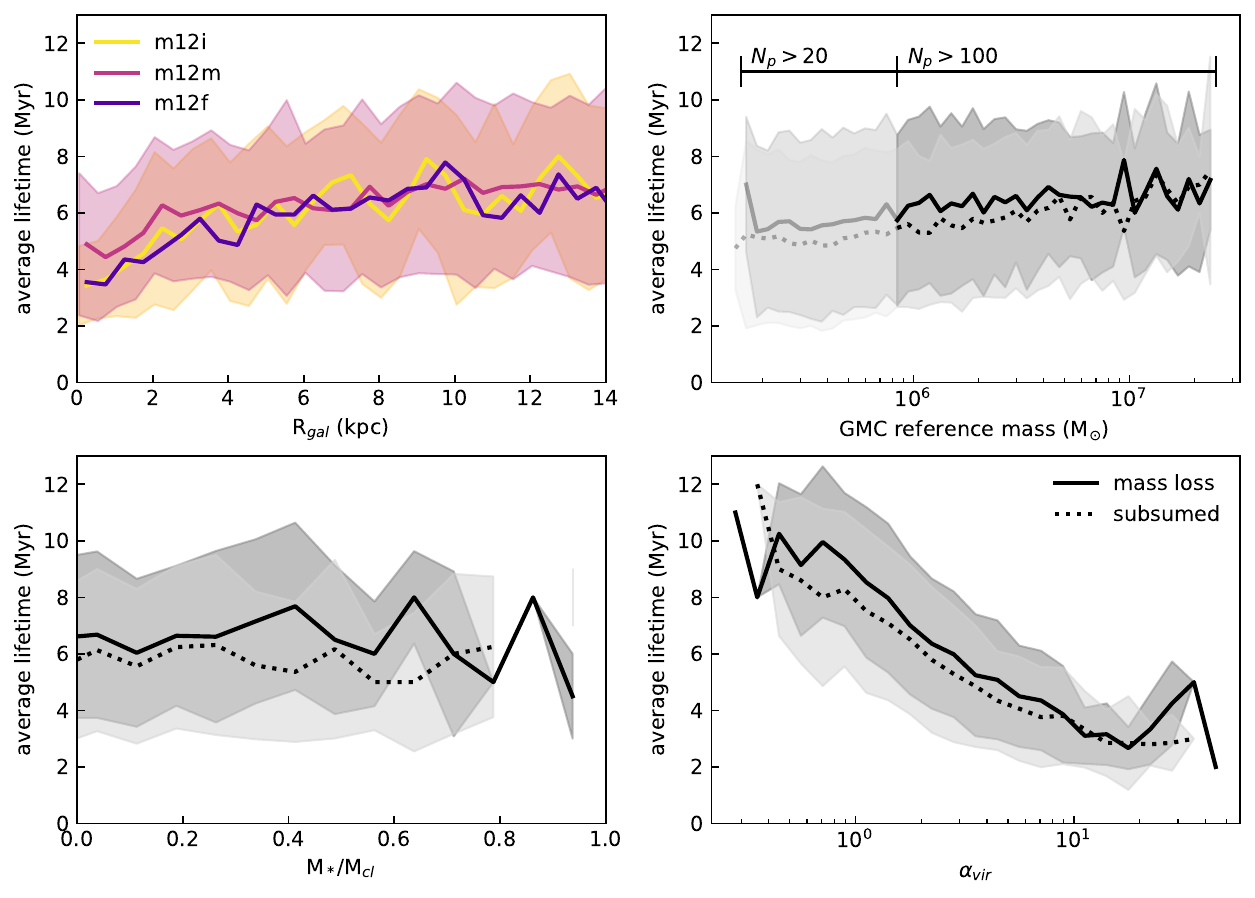}
    \caption{The dependence of GMC lifetimes on GMC and environmental properties. Each of the panels shows all clouds with more than 100 gas elements to examine the best resolved population. In all gray-scale panels, solid lines denote the clouds that died via mass loss and dotted lines denote clouds that have died by being subsumed into a larger cloud. \textit{Top Left:} The dependence of GMC lifetime on galacto-centric radius, subdividing the sample by the host galaxy. \textit{Top Right:} The dependence of GMC lifetime on the GMC mass at its reference snapshot. Here, we show clouds down to 20 particles to extend the mass range. \textit{Bottom Left:} The dependence GMC lifetime on the fraction of the GMC mass that is converted to stars over its lifetime. \textit{Bottom Right:} The dependence of GMC lifetime on the virial parameter of GMC as measured at the reference time. This is the only quantity that shows any significant dependence. We find that GMCs with short lifetimes are biased to having larger virial parameters or being less gravitationally bound. }
    \label{fig:boxplot}
\end{figure*}

\begin{table}
\centering
	\caption{Mean and standard deviation of GMC lifetimes across our 3 cosmological simulations, for all GMCs, those that form stars, and the different destruction mechanisms. $N_p$ indicates the minimum number of gas elements, effectively a mass threshold.}
	\begin{threeparttable}
	\begin{tabular}{cccccc}
		\hline
		 && \multicolumn{2}{c}{all GMCs} &  \multicolumn{2}{c}{formed stars} \\
		 & $N_p$ & $\langle l \rangle$ (Myr) & $\sigma_l$ (Myr) & $\langle l \rangle$ (Myr)& $\sigma_l$ (Myr)\\
		 \hline
		{\tt m12i} & 20 & 5.77 & 3.14 & 5.90 & 3.00 \\
		 & 100 & 6.36 & 2.82 & 6.32 & 2.83\\
		 \hline
		 {\tt m12m} & 20 & 5.50 & 3.04 & 5.82 & 3.11 \\
		 & 100 & 6.59& 2.98 & 6.69 & 2.95 \\
		 \hline
		{\tt m12f} & 20 & 5.39 & 2.94 & 5.50 & 2.90\\
		 & 100 & 6.05 & 2.57 & 6.02 & 2.44\\
		 \hline
		 && \multicolumn{2}{c}{mass loss GMCs} &  \multicolumn{2}{c}{subsumed GMCs} \\
		 & $N_p$ & $\langle l \rangle$ (Myr) & $\sigma_l$ (Myr) & $\langle l \rangle$ (Myr)& $\sigma_l$ (Myr)\\
		 \hline
		{\tt m12i} & 20 & 6.16 & 3.14 & 5.58 & 3.12 \\
		 & 100 & 6.48 & 2.80 & 6.26 & 2.83\\
		 \hline
		 {\tt m12m} & 20 & 6.01 & 3.13 & 5.27 & 2.97\\
		 & 100 & 7.03 & 2.96 & 6.29 & 2.95 \\
		 \hline
		{\tt m12f} & 20 & 5.86 & 3.04 & 5.19 & 2.88\\
		 & 100 & 6.27 & 2.60 & 5.87 & 2.53\\
		 \hline

	\end{tabular}
	\end{threeparttable}
	\label{tab:life}
\end{table}

\subsection{The variation of GMC lifetimes with environment and GMC properties}

The distributions of GMC lifetimes in Figure~\ref{fig:life} show little galaxy-to-galaxy variation. We now ask how these lifetimes depend on GMC properties and the environmental conditions in the galactic disk. Figure~\ref{fig:boxplot} shows how GMC lifetimes depend on galactocentric radius (top left), the GMC mass at the reference time (top right), the fraction of mass in a cloud that is converted to stars during its lifetime (bottom left), and the GMC's virial parameter at the reference time (bottom right). Only for the case of galactocentric radius do we separate the GMCs by their host galaxy, to account for any possible differences in the host galaxies. In the remaining cases we compile all 3 host galaxies as a single population, and we separate by the destruction mechanism: whether the cloud died via mass loss or by being subsumed. In all of the plots we include only our most-resolved population of GMCs, those with $N_p\!>\!100$. The only exception is the plot showing the GMC reference mass (top right), here we include the entire sample of clouds to help extend the mass range considered to discern if a trend is present.

We find that the GMC lifetimes increase slightly from the center to the edge of the galaxy, by $\sim2$ Myr on average. This is consistent with the trend of the galactic free-fall time with galactocentric radius, which similarly increases by $\sim2$ Myr from the inner to outer-most parts of the galaxies. Here, the free-fall time, $t_{ff}=\sqrt(3/32\pi G \rho)$, is measured on 100 pc apertures centred on the GMCs. We similarly see a slight increase of the lifetime with GMC reference mass, which is consistent with trends seen in other work \citep[e.g.][]{oklopcic2017, hopkins2012}. Specifically, we find an increase in the mean cloud lifetime of 3 Myr, and an increase in the median of 4 Myr, across the plotted mass range. We conversely find no trend with the amount of the GMC that is converted to stars. Further, there is only a weak offset between the populations clouds destroyed via mass loss versus those that are subsumed, but the dependence with these properties is similar for the two populations regardless.

The quantity for which we see a strong dependence is the cloud virial parameter, $\alpha_{vir}$, as measured at the reference time. As noted previously, our GMCs display a large range of virial parameters (boundedness), with a mean of 3.1 and median of 2.2 for the full population of stacked GMCs. In Figure~\ref{fig:boxplot} clouds with large virial parameters, that is, those that are loosely gravitationally bound, have shorter lifetimes. As expected, these unbound clouds disrupt on a dynamical time.

\subsection{Observational methods of determining GMC lifetimes}
To make meaningful comparisons to observational inferences of GMC lifetimes, one must understand the methods and assumptions used. In this work, we measure GMC lifetimes via the length of time that gas remains in an identified overdense structure. Many different definitions of GMC lifetimes have been used, ranging from the emission lifetime of CO to the lifetime of H$_2$ molecules themselves, and it is important to understand these differences as we make comparisons to observationally determined lifetimes. We will pursue similar observationally based metrics of cloud lifetimes and comparisons of these in future works.  However, for context, we review and compare to findings from a selection of other methods.

One approach is to infer lifetimes based on the positions of GMCs along HI filaments. In M33, for example, most GMCs still are associated with their HI filaments, which suggests that they do not live long enough to drift across/off the filament: using this method, \citet{engargiola2003} infer an upper limit of $10-20$ Myr on GMC lifetimes.

Another approach is to assume that GMCs go through different evolutionary states and classify them accordingly: this method requires correlating catalogs for GMCs, HII regions, and young stellar objects. Using this methodology inferred multiple studies have inferred that GMC lifetimes are $20-40$ Myr \citep[e.g.][]{kawamura2009, miura2012}.

\citet{kruijssen2014} and \citet{kruijssen2018} use a statistical method to measure GMC lifetimes. This is enabled using the uncertainty principle for star formation, which assumes that there is a correlation between GMCs and star-forming regions at different scales. \citet{kruijssen2019} applied this method to NGC 300, inferring GMC lifetimes of $\sim 10$ Myr \citep{kruijssen2019}. \cite{chevance2020} greatly extended the sample and used the above mentioned statistical method to measure GMC lifetimes in a further nine galaxies in the PHANGS sample, where they find short lifetimes, between 9 and 30 Myr depending on the chosen galaxy. \cite{schinnerer2019b} employ a simplified version of this approach to measure the cold gas timescale for a selection of galaxies in the PHANGS sample. Using this approach they recover a GMC lifetime of between 10-15 Myr.

In summary, observational inferences suggest GMC lifetimes of $10-40$ Myr. In comparison, our cosmological simulations of MW/M31-like galaxies, which resolve massive GMCs, have mean lifetimes of $\sim6$ Myr and maximum lifetimes of $\sim20$ Myr.
While our measurements favour shorter lifetimes, they are broadly consistent with both numerical \citep[e.g.][]{dobbs2013, grisdale2019} and observational studies (see above). It is worth noting that these initial results are sensitive to the method used. Specifically, the mass fraction required for cloud survival plays a large role. Lowering this limit can increase the cloud lifetime by at most a factor of 2.

An obstacle to drawing meaningful conclusions from theoretical studies of cloud formation is to decipher how we should compare findings to observational metrics. In upcoming work we will provide an in depth analysis of how to interpret theoretical cloud lifetimes in comparison to H$_2$ and CO.

\section{Summary and Conclusions} \label{sec:end}
For the first time we have measured the lifetimes of GMCs in cosmological simulations. We find average GMC lifetimes between 5 and 7 Myr, with few clouds surviving past 20 Myr. We find little variation in the distribution of GMC lifetimes across our 3 simulated galaxies ({\tt m12m}, {\tt m12i}, and {\tt m12f}). We find limited dependence of the lifetime on GMC mass, although we can resolve only the most massive clouds, with total gas mass $\gtrsim 10^5$ M$_\odot$, and we see little dependence on the star formation activity in the cloud. We do find weak dependence of GMC lifetime on galactic environment, with a small increase in the cloud lifetime with increasing galactocentric radius. We do find that GMC lifetimes depend strongly on the cloud virial parameter, with less gravitationally bound clouds exhibiting shorter lifetimes.

The GMC lifetime may be set, in part, by transient compression of gas as it moves through the spiral arms. We plan to explore whether the forcing of structures on timescales shorter than the orbital is contributing to our overall cloud lifetimes. Forthcoming papers will focus on better understanding the connection between GMCs and star formation, including how cloud lifetimes vary as a function of environment and cloud evolutionary history.

\section*{Acknowledgements}

The authors thank the scientific editor, Joop Schaye, and the anonymous referee whose insight greatly improved the work presented here.  This research made use of Astropy,\footnote{http://www.astropy.org} a community-developed core Python package for Astronomy \citep{astropy2013, astropy2018}. SB, AW, and MB were supported by a Hellman Fellowship, the Heising-Simons Foundation, and NASA, through ATP grant 80NSSC18K1097 and HST grants GO-14734 and AR-15057 from STScI. SRL was supported by NASA through Hubble Fellowship grant \#HST-JF2-51395.001-A awarded by STScI. NM was supported by NSERC. PFH was supported by NSF grants 1715847 \&\ 1911233, CAREER grant 1455342, and NASA grants 80NSSC18K0562 \& JPL 1589742. DG was supported by the Harlan J. Smith McDonald Observatory Postdoctoral Fellowship. CAFG was supported by NSF grants AST-1517491, AST-1715216, CAREER award AST-1652522, by NASA grant 17-ATP17-0067, by STScI grant HST-GO-14681.011, and by a Cottrell Scholar Award from the Research Corporation for Science Advancement.
We ran these simulations and analysis using allocations from: XSEDE AST130039 \& AST140064; Blue Waters PRAC NSF.1455342 supported by NSF; and NASA HEC SMD-16-7324, SMD-16-7592, SMD-17-1289, SMD-17-1375, SMD-18-2189.

\section*{Data Availability}
Single simulation snapshots at $z$=0 are available for {\tt m12i}, {\tt m12f}, and {\tt m12m} at ananke.hub.yt. Our analysis made use of the publicly available python packages https://bitbucket.org/awetzel/gizmo\_analysis and https://bitbucket.org/awetzel/utilities.




\bibliographystyle{mnras}
\bibliography{FIREgmcs} 

\appendix

\section{Impact of Mass Fraction Choice} \label{sec:app}

\begin{figure*}
    \centering
    \includegraphics[width=0.95\textwidth]{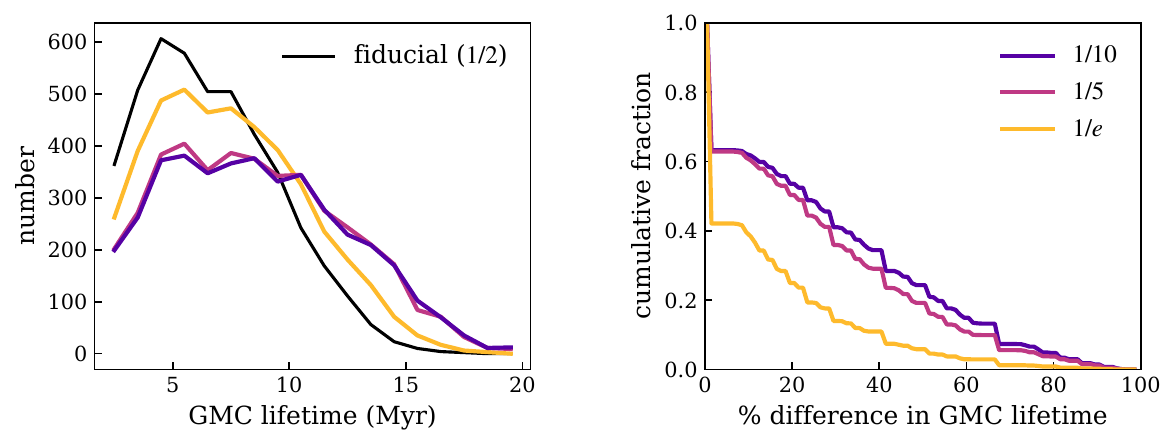}
    \caption{Quantifying the change in GMC lifetime based on the mass fraction threshold for destruction. If clouds are not subsumed into a larger cloud, destruction occurs based on the fraction of the reference cloud mass remaining at any given time. In our fiducial case, a cloud's life ends if its mass drops below half of its reference mass. Here we show the impact of changing this threshold to fractions of $1/e$, $1/5$, and $1/10$. \textit{Left:} Distribution of lifetimes for each of the cases, stacking all three hosts. \textit{Right:} A reverse cumulative histogram of the percent difference in cloud lifetime for each mass fraction threshold. To assess this difference, clouds in the fiducial, $1/2$, as matched to clouds in each of the lower mass fraction cases: this produces a direct cloud-to-cloud comparison on the impact of changing the mass fraction threshold.  }
    \label{fig:apmass}
\end{figure*}

In the work presented here we have tracked the evolution of GMCs as described in section \ref{sec:track}. In this algorithm, a cloud's lifetime ends either when it has lost a certain fraction of its mass or when it is subsumed by another cloud. The use of a mass fraction, or mass loss threshold, of course necessitates the selection of a cutoff point. In this appendix we outline the consequences of choosing such a threshold, and the dependence that our GMC lifetimes have on this selection.

We choose $1/2$ to be our fiducial mass fraction as it allows this work to be placed in the context of other theoretical measurements, where $1/2$ is a common choice \citep[see for e.g.][]{dobbs2013}. Figure~\ref{fig:apmass} shows the impact of decreasing this threshold to $1/e$, $1/5$ and $1/10$ of the reference GMC mass. As expected, decreasing this threshold extends the GMC lifetime.

The left panel of Figure~\ref{fig:apmass} shows the lifetime distribution for the different mass-fraction thresholds. In this case each line represents a stacking of all 3 host galaxies. Here we see the overall impact of the threshold illustrated, relaxing the minimum mass fraction extends the cloud lifetimes and makes the distributions wider. Generally though, the change to the peak lifetime for the distributions is small, on the order of 1-2 Myr. Further, we see this effect taper off: convergence appears to occur between a fraction of $1/5$ and $1/10$.

The right panel of Figure~\ref{fig:apmass} shows a reverse cumulative histogram of the percent difference in GMC lifetime, in comparison to the fiducial mass fraction threshold ($1/2$). This in essence matches GMCs across different tracking cases and quantifies how much their lifetime is extended by relaxing the mass fraction, in essence allowing them to be less massive relative to their reference mass and still remain alive. As to be expected, there are of course differences as this threshold is relaxed. However, those differences are small for the majority of GMCs; even a difference of 25\% corresponds to at most 2 Myr, but on average less than that.


\bsp	
\label{lastpage}
\end{document}